\documentclass[prl,showpacs,byrevtex,floatfix,superscriptaddress,twocolumn]{revtex4}
\usepackage{graphicx}
\usepackage{float}
\usepackage{bm}
\begin{document}

\bibliographystyle{apsrev}

\preprint{Draft version, not for distribution}

%
%
\title[boson]{Signatures of electron-boson coupling in half-metallic
ferromagnet Mn$_5$Ge$_3$: \\ study of electron self-energy
$\Sigma(\omega)$ obtained from infrared spectroscopy}
%
%
\author{S. V. Dordevic}
\email{dsasa@uakron.edu}%
\affiliation{Department of Physics, The University of Akron,
Akron, Ohio 44325 USA}%
\author{N. Stojilovic}
\affiliation{Physics Department, John Carroll University,
University Heights, OH 44118 USA}
\author{L. W. Kohlman}
\affiliation{Department of Physics, The University of Akron,
Akron, Ohio 44325 USA}
\author{Rongwei Hu}
\affiliation{Condensed Matter Physics and Materials Science
Department, Brookhaven National Laboratory, Upton, New York 11973, USA}
\affiliation{Department of Physics, Brown University, Providence, RI 02912, USA}
\author{C. Petrovic}
\affiliation{Condensed Matter Physics and Materials Science
Department, Brookhaven National Laboratory, Upton, New York 11973, USA}

\date{\today}

%
%
\begin{abstract}
We report results of our infrared and optical spectroscopy study
of a half-metallic ferromagnet Mn$_5$Ge$_3$. This compound is
currently being investigated as a potential injector of spin
polarized currents into germanium. Infrared measurements have been
performed over a broad frequency (50 - 50000~cm$^{-1}$) and
temperature (10 - 300 K) range. From the complex optical
conductivity $\sigma(\omega)$ we extract the electron self-energy
$\Sigma(\omega)$. The calculation of $\Sigma(\omega)$ is based on
novel numerical algorithms for solution of systems of non-linear
equations. The obtained self-energy provides a new insight into electron
correlations in Mn$_5$Ge$_3$. In particular, it reveals that charge
carriers may be coupled to bosonic modes, possibly of magnetic origin.
\end{abstract}

%
%
%
%
%

\maketitle

\begin{figure}[t]
\vspace*{-1.0cm}%
\centerline{\includegraphics[width=9.5cm]{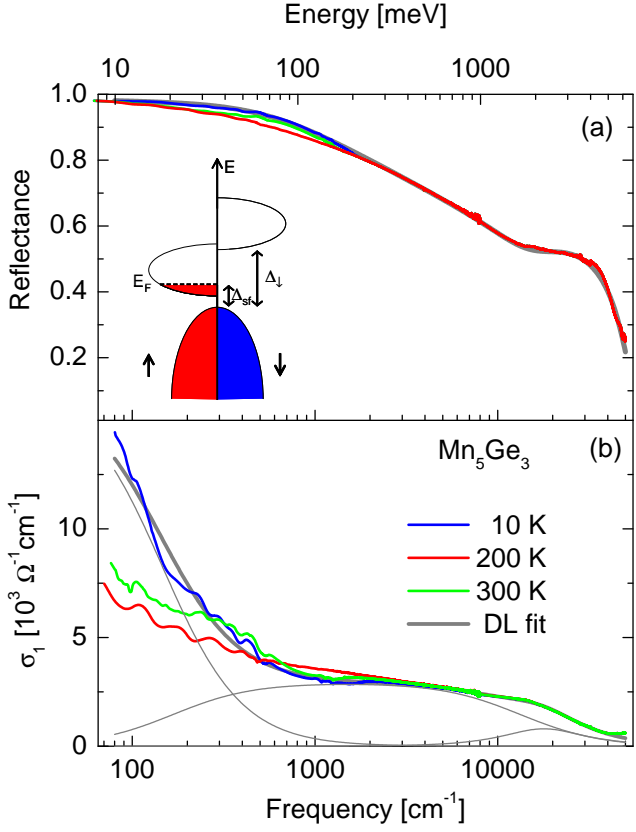}}%
\vspace*{-1.0cm}%
\caption{(Color online). (a) Reflectance of Mn$_5$Ge$_3$ at
several selected temperatures between 10 and 290~K. The temperature
dependence is limited mostly to the region below about
1500~cm$^{-1}$. The inset represents a schematic band structure
for a half-metallic ferromagnet, the bands for spin-up and
spin-down electrons plotted separately \cite{irkhin94,pickett01,katsnelson08}.
The energy gap for minority spin electrons is $\Delta_{\downarrow}$,
and the spin-flip gap is $\Delta_{sf}$. (b) The optical conductivity
$\sigma_1(\omega)$ obtained from reflectance using KK
transformation. Also shown with gray lines are the total DL fit at
10~K, as well as the three individual components of the fit.}
\vspace*{-0.5cm}%
\label{fig:ref}
\end{figure}

Half-metallic ferromagnetism (HMFM) was first discovered in
half--Heusler alloy NiMnSb \cite{degroot83}. Band structure
calculations showed that all electrons at the Fermi level had
the same spin orientation, resulting in a spin-polarized
ferromagnetic state
\cite{irkhin94,pickett01,katsnelson08}. This was in stark contrast
with conventional metals like gold or copper, in which there is an
equal population of spin-up and spin-down electrons, resulting in
a non-magnetic ground state. The inset of Fig.~\ref{fig:ref}
displays schematic one-electron band structure of HMFM. In an ideal
case, the number of electrons at the Fermi level with one spin
orientation is zero (minority spin electrons), i.e. all
electrons have the opposite spin orientation (majority spin
electrons). In real materials there may be electrons with both
spin polarizations at the Fermi level. Their number is
characterized through the so-called spin polarization
$P=(N_{\uparrow}-N_{\downarrow})/(N_{\uparrow}+N_{\downarrow})$
\cite{mazin99}.

In this work we have used infrared (IR) spectroscopy to study
Mn$_5$Ge$_3$, an intermetallic compound shown to exhibit HMFM with
a Curie temperature of T$_C \simeq$~293~K \cite{fontaine62}.
Mn$_5$Ge$_3$ crystallizes in a hexagonal crystal structure, space
group P63/mcm, with lattice parameters a = 7.191~$\AA$ and c =
5.054~$\AA$. The anisotropic crystal structure results in slightly
anisotropic transport and thermodynamic properties
\cite{petrovic09,picozzi04}. Recent theoretical calculations and
spin-sensitive point contact Andreev refection measurements have
shown that Mn$_5$Ge$_3$ has a substantial spin polarization in the
range 43-54 $\%$ \cite{panguluri05}. In addition to its unusual
properties from the fundamental point of view, Mn$_5$Ge$_3$ has
also attracted attention as a potential injector of spin-polarized
currents into germanium (Ge) \cite{zeng03,panguluri05}. Thin films
of Mn$_5$Ge$_3$ have been successfully grown on Ge(111) using
solid phase epitaxy \cite{zeng03}. This important technological
advance opens the door for integration of Mn$_5$Ge$_3$ with the
existing Ge technologies and possible spintronics applications.


Reflection measurements on single crystal Mn$_5$Ge$_3$ have been
performed over a broad frequency range from about 50~cm$^{-1}$ to
50000~cm$^{-1}$ (6~meV -- 6.2~eV) and at temperatures from 10~K up to room
temperature. An overfilling technique was used to obtain the absolute
values of reflectance from a sample with surface area of
approximately 1~mm $\times$ 1~mm \cite{homes93}. From the measured
reflectance R($\omega$), complex optical conductivity
$\sigma(\omega)=\sigma_1(\omega)+i\sigma_2(\omega)$ was obtained
through Kramer-Kroning (KK) transformation
\cite{basov05,dordevic06}.


Figure \ref{fig:ref} displays raw reflectance data R($\omega$)
and the real part of optical conductivity $\sigma_1(\omega)$.
The overall behavior of both optical functions is
metallic, however there are some features that distinguish it
from conventional metals \cite{dressel-book}. One can identify
three contributions to $\sigma_1(\omega)$: i) a Drude-like peak at
zero frequency which narrows as temperature decreases, ii) a
region of relatively constant conductivity (between about 1000 and
10000~cm$^{-1}$) and iii) a region of interband transitions above
10000~cm$^{-1}$. We also note that similar three contributions
to $\sigma_1(\omega)$ were observed in some other HMFL, such as CrO$_2$
\cite{singley99} and in NiMnSb \cite{mancoff99}.


To gain further insight into electronic properties of Mn$_5$Ge$_3$
we employed a standard Drude-Lorentz (DL) model
\cite{dressel-book,basov05,dordevic06}.The minimal model to achieve
a good fit consisted of a Drude and two Lorentzian modes, centered at
around 1480~cm$^{-1}$ and 18000~cm$^{-1}$. The total fit, as well as
the three individual contributions, are shown in Fig.~\ref{fig:ref}(b)
with gray lines. The oscillator at 18000~cm$^{-1}$ simulates the effect
of interband transitions. On the other hand, the oscillator at
1480~cm$^{-1}$ seems to indicate the presence of low-lying transition
and we tentatively assign it to excitations across the spin-flip gap
($\Delta_{sf}$ in Fig.~\ref{fig:ref}). We will come back to this
important point below.

\begin{figure}[t]
\vspace*{-1.0cm}%
\centerline{\includegraphics[width=9.5cm]{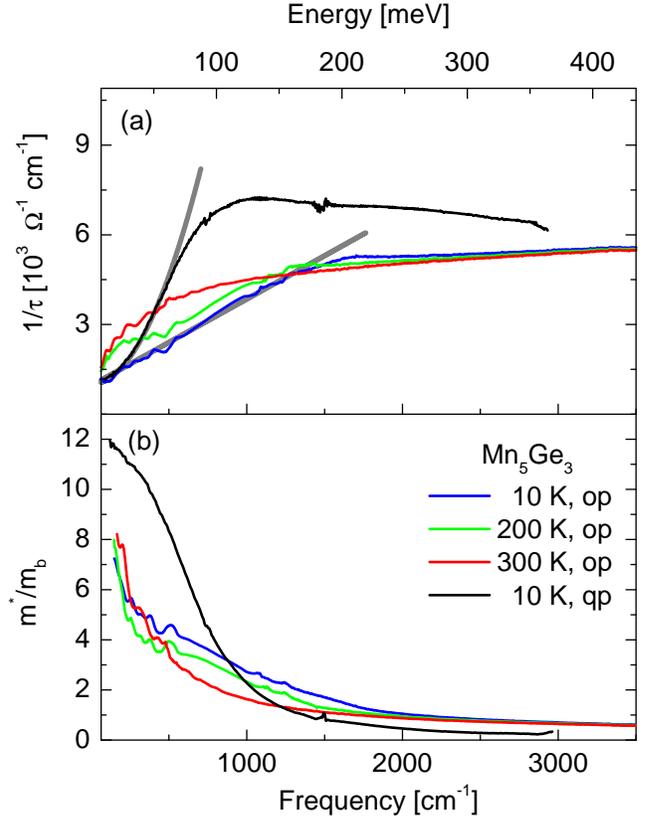}}%
\vspace*{-1.0cm}%
\caption{(Color online). Optical scattering rate
$1/\tau(\omega)^{op}$ and effective mass $m^*(\omega)^{op}/m_b$
obtained from Eqs.~\ref{eq:tau} and \ref{eq:mass}. Also shown are
the quasiparticle scattering rate $1/\tau(\omega)^{qp}$ and
effective mass $m^*(\omega)^{qp}/m_b$ at 10~K, as defined in the text.}%
\vspace*{-0.5cm}%
\label{fig:tau}
\end{figure}

In the one-component approach, one assumes that only a single type
of carriers are present in the system, but their scattering rate
acquires frequency dependence \cite{basov05,dordevic06,dressel-book}.
Within the so-called "extended"
Drude model one calculates the {\it optical} scattering rate
$1/\tau(\omega)^{op}$ and effective mass $m^*(\omega)^{op}/m_b$
from the optical conductivity $\sigma(\omega)$ as:

\begin{equation}
\frac{1}{\tau(\omega)^{op}}=\frac{\omega_{p}^{2}}{4 \pi} \Re \Big[
\frac{1} {\sigma(\omega)} \Big]
\label{eq:tau}%
\end{equation}

\begin{equation}
\frac{m^{*}(\omega)^{op}}{m_b} = \frac{\omega_{p}^{2}}{4 \pi} \Im
\Big[ \frac{1} {\sigma(\omega)} \Big] \frac{1}{\omega}
\label{eq:mass}%
\end{equation}
where the plasma frequency $\omega_p^2=4 \pi e^2 n/m_b$ ($n$ is the
carrier density and $m_b$ their band mass)is
estimated from the integration of $\sigma_{1}(\omega)$ up to the
frequency of the onset of interband absorption (the cut-off
frequency). By integrating the optical conductivity up to
10000~cm$^{-1}$ we get $\omega_p$=~33000~cm$^{-1}$ (4.1~eV) \cite{comm-plasma}.
However the value of plasma frequency obtained this way
is ambiguous, as it depends on the choice of the cut-off frequency,
i.e. the separation of intra- and inter-band contributions.
The results of extended Drude analysis are shown in
Fig.~\ref{fig:tau}. The optical scattering rate 1/$\tau(\omega)^{op}$
displays almost linear frequency dependence, and characteristic
suppression below about 1500~cm$^{-1}$. This is very close
to the frequency of the low-lying transitions obtained from
DL fits. We also point out that similar suppression of scattering
rate was observed in the spectra of HMFM CrO$_2$
\cite{singley99} and NiMnSb \cite{mancoff99}.


To elucidate this behavior we employ an alternative approach of
quantifying correlation effects: we calculate the electron
self-energy $\Sigma(\omega)$ directly from $\sigma(\omega)$.
Similar procedure was used before in studies of superconductors
\cite{gimm02,gimm03}. Electron self-energy is a physical quantity
of significant interest, as it contains important information
about electron interactions in a solid. In general, it is a
complex function of frequency, $\Sigma(\omega) = \Sigma_1(\omega) +
i \Sigma_2(\omega)$, where its real part, $\Sigma_1(\omega)$, is
related to quasiparticle energy renormalizaiton and the
imaginary part $\Sigma_2(\omega)$ corresponds to quasiparticle
life-time.

Hwang et al. introduced \cite{hwang04,op-qp-comm} {\it optical} self-energy
$\Sigma^{op}(\omega)$, which has been used frequently in the analysis
of IR data \cite{hwang08,heumen08,basov09}. $\Sigma^{op}(\omega)$ is
defined as:

\begin{equation}
\sigma(\omega)=\frac{\frac{i \omega_p^2}{4 \pi}}{\omega - 2
\Sigma^{op}(\omega)}
\label{eq:timusk}%
\end{equation}
and it can be calculated from $\sigma(\omega)$ using an analytic
transformation. It is easy to show that $1/\tau(\omega)^{op} = - 2
\Sigma_2^{op}(\omega)$ and $\omega(m^*(\omega)^{op}/m_b-1) = -2
\Sigma_1^{op}(\omega)$.
Here we calculate the electron self-energy directly from its relation
to $\sigma(\omega)$ at T=~0~K \cite{allen71,allen04,norman06}:

\begin{equation}
\sigma(\omega)=\frac{i\omega_p^2}{4 \pi \omega} \int_{0}^{\omega}
\frac{d\omega'}{\omega-\Sigma
(\omega'+\omega)+\Sigma^{\ast}(\omega')}.
\label{eq:se2}%
\end{equation}
This expression reduces to Eq.~\ref{eq:timusk} as the first order
approximation in $\omega$ \cite{allen04}. When discretized,
Eq.~\ref{eq:se2} becomes:

\begin{equation}
\sigma_k=\frac{i \omega_p^2}{4 \pi \omega_k}
\sum_{j=1}^k \frac{\Delta \omega}{\omega_k - \Sigma_{j+k} + \Sigma_j^{*}},
\label{eq:se3}%
\end{equation}
where $\sigma_k=\sigma(\omega_k)$, $\Sigma_k=\Sigma(\omega_k)$,
k=1,... N, and $N$ is the number of (equidistant) points at which
we perform the inversion. Eq.~\ref{eq:se3} effectively represents 
a system of 2N {\it non-linear} equations with 2N unknowns
($\Sigma_{1k}$ and $\Sigma_{2k}$), but since $\Sigma(\omega)$ is a
response function, its real and imaginary parts are related
through a KK transformation. Thus the problem reduces to a system
of N equations with N unknowns (for example $\Sigma_{2k}$).
However it is an inverse problem and is also ill-posed.

Ill-posed inverse problems can be solved using, for example, the
Levenberg–-Marquardt algorithm \cite{nr}. To suppress numerical
instabilities, the problem must first be regularized \cite{nr}. We
have employed first order Tikhonov regularziation, which seems to
give the best results. The solution of the system starts from an
initial guess and proceeds iteratively, until it has converged to
within the required accuracy. An interesting choice for the
starting point of iteration is $\Sigma^{op}(\omega)$ from
Eq.~\ref{eq:timusk}. One also must ensure that the system converges 
to the same solution from several different starting points.


\begin{figure}[t]
\vspace*{-1.0cm}%
\centerline{\includegraphics[width=9.5cm]{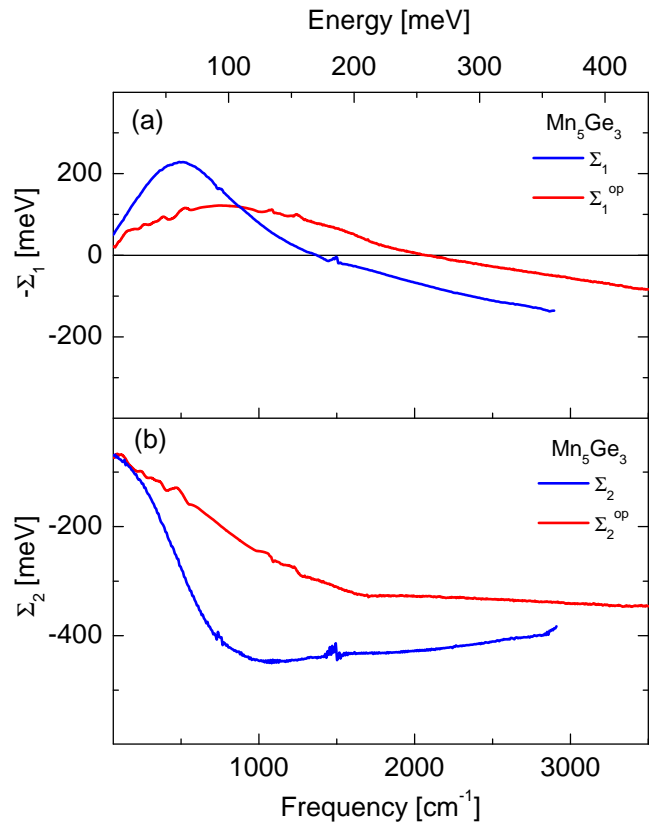}}%
\vspace*{-1.0cm}%
\caption{(Color online). Real and imaginary parts of self-energy
$\Sigma(\omega)$ obtained from the numerical inversion of
Eq.~\ref{eq:se3}. Also shown are the real and imaginary parts of
optical self-energy $\Sigma^{op}(\omega)$ calculated from
Eq.~\ref{eq:timusk}.}%
\vspace*{-0.5cm}%
\label{fig:se}
\end{figure}


Figure \ref{fig:se} displays the results of inversion calculations
for Mn$_5$Ge$_3$ at 10 K, as well as the optical self-energy
$\Sigma^{op}(\omega)$ from Eq.~\ref{eq:timusk}. As can be seen,
there are some differences between $\Sigma(\omega)$ and
$\Sigma^{op}(\omega)$. By analogy with the optical scattering
rate 1/$\tau(\omega)^{op}$ (Eq.~\ref{eq:tau}) and effective mass
$m^*(\omega)^{op}/m_b$ (Eq.~\ref{eq:mass}) we introduce {\it quasiparticle}
scattering rate as 1/$\tau(\omega)^{qp} = - 2 \Sigma_2(\omega)$
and {\it quasiparticle} effective mass as
$\omega(m^*(\omega)^{qp}/m_b-1) = -2 \Sigma_1(\omega)$. These two
new quantities are shown in Fig.~\ref{fig:tau}, and they also
differ from their optical counterparts. Most notably, the lower energy
suppression of scattering rate follows a power-law behavior:
1/$\tau(\omega)^{qp}\simeq \omega^2$, whereas
1/$\tau(\omega)^{op}\simeq \omega$ (gray lines in Fig.~\ref{fig:tau}).
These important differences warrant further theoretical considerations.


As mentioned above, the low frequency suppression of scattering rate
observed in Mn$_5$Ge$_3$ was also found in CrO$_2$ and NiMnSb, and
appears to be generic to all HMFM. The suppression 1/$\tau(\omega)$
can be explained as a consequence of HMFM band structure (inset of
Fig.~\ref{fig:ref}). Namely, in the one-electron picture, the low-energy
minority spin excitations (spin-flip processes) are forbidden because
of the lack of available states \cite{katsnelson08}. This scenario
seems to work well in CrO$_2$ where the scattering rate drops to very
low values below about 500~cm$^{-1}$, which was assigned to to
spin-flip gap \cite{singley99}. In Mn$_5$Ge$_3$ the
scattering rate does not drop to zero, but is, instead, suppressed
following a power-law dependence. This behavior can, in principle,
be explained as due to peculiarities of the density of states
\cite{sharapov05}. We propose an alternative explanation in terms
of the low-energy scattering channels. They require one to go
beyond the one-electron picture \cite{katsnelson08}. Namely, when
electron correlations are taken into account, the low-energy spin-flip
excitations become possible because of a coupling of majority spin
electrons and virtual magnons \cite{katsnelson08}. This coupling is
believed to be of prime importance for transport processes
in HMFM (and ferromagnets in general).

\begin{figure}[t]
\vspace*{2.0cm}%
\centerline{\includegraphics[width=9.5cm]{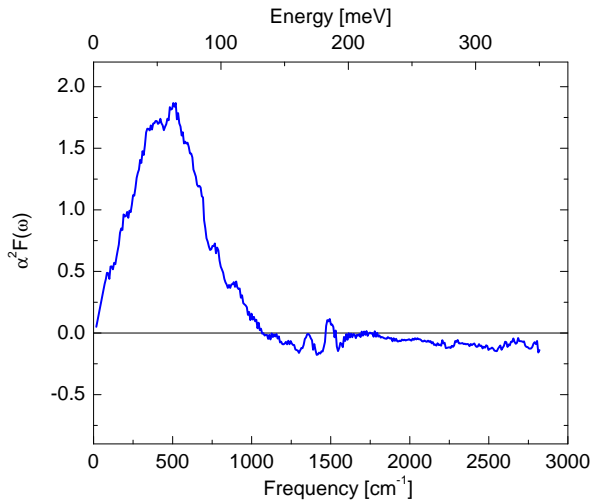}}%
\vspace*{-2.5cm}%
\caption{(Color online). The electron-boson spectral function
$\alpha^2F(\omega)$ of Mn$_5$Ge$_3$ calculated from $\Sigma_2(\omega)$.}%
\vspace*{-0.5cm}%
\label{fig:alpha2F}
\end{figure}

Coupling of charge and bosonic degrees of freedom has been know to be
important in solids. In optical functions this coupling shows up as a
suppression of scattering rate \cite{basov05,dordevic05,dordevic06}.
A quantity that contains all of the information about such coupling is
the electron-boson spectral function $\alpha^2F(\omega)$, which can
be obtained from $\Sigma(\omega)$ as $-\pi \alpha^2F(\omega) =
d\Sigma_2(\omega)/d\omega$ \cite{norman06}. $\alpha^2F(\omega)$
calculated this way is shown in Fig.~\ref{fig:alpha2F} and, as could
be anticipated from $1/\tau(\omega)^{qp}$, it is characterized by a
pronounced peak centered at about 500~cm$^{-1}$ (60~meV). This peak
indicates that charge carriers may be coupled to bosonic modes,
such as magnons. Note however that in Mn$_5$Ge$_3$ the
spin-polarization is only partial \cite{panguluri05}, which implies
that other scattering channels (such as phonons) are also possible.
Further experimental tests, such as magneto-optical measurements,
will help elucidate if this scenario applies to Mn$_5$Ge$_3$,
as well as the contributions of various scattering channels.


High Curie temperature, substantial degree of spin polarization, and a
good lattice match with germanium and silicon make Mn$_5$Ge$_3$ a
very promising candidate for an injector of spin-polarized currents.
Recently discovered \cite{clc07} simultaneous enhancement of both
spin polarization and Curie temperature in Fe--doped Mn$_5$Ge$_3$
above room temperature opens new and exciting possibilities for
controlled spin injection in functional spintronics devices. In this
work we used IR spectroscopy to study electronic properties of
Mn$_5$Ge$_3$. Our results reveal the spin-flip gap of approximately
1500~cm$^{-1}$ (186~meV). The inversion procedure described
allows access to electron self-energy from IR, and more direct (and
quantitative) comparisons with photoemission spectroscopy. Both
optical and quasiparticle scattering rates reveal a power-law
behavior (albeit with different exponents), which might signal
coupling of charge carriers to bosonic modes, possibly of magnetic origin.


We thank K.S. Burch for bringing Ref.~\cite{gimm02} to our
attention and D.N. Basov, C.C. Homes, V.Yu. Irkhin
and T. Timusk for critical reading of the manuscript. Special
thanks to R.D. Ramsier for the use of his equipment.
Part of this research was carried out at the Brookhaven
National Laboratory, which is operated for the U.S. Department of
Energy by Brookhaven Science Associates (DE-AC02-98CH10886).

%
%

\begin{references}
\bibitem{comm-plasma} This value cannot be directly compared with
these obtained from numerical calculations \cite{picozzi04} because
IR cannot resolve the two spin directions.
\bibitem{op-qp-comm} Hwang at al. \cite{hwang07} have also introduced a quantity
$\Sigma^{op-qp} \equiv d[\omega \Sigma^{op}(\omega)]/d\omega$.
\end{references}

\end{document}